\documentclass[twocolumn,prl,aps,showpacs]{revtex4}

\usepackage{amssymb}
\usepackage{amsmath}
\usepackage{graphicx}
\usepackage{dcolumn}
\usepackage{bm}

\begin{document}

\title{Ultrafast initialization and QND-readout of a spin qubit via
    control of nanodot-vacuum coupling}

\author{Ren-Bao Liu}
\author{Wang Yao}
\author{L. J. Sham}

\affiliation{Department of Physics, University of California San
Diego, La Jolla, California 92093-0319}

\date{\today}

\begin{abstract}
Ultrafast initialization enables fault-tolerant processing of quantum
information while QND readout enables scalable
quantum computation. By spatially assembling photon resonators and
wave-guides around an n-doped nanodot and by temporally designing
optical pump pulses, an efficient quantum pathway can be
established from an electron spin to a charged exciton to a cavity
photon and finally to a flying photon in the waveguide. Such
control
of vacuum-nanodot coupling can be exploited for
ultrafast initialization and QND readout of the spin, which are
particularly compatible with the optically driven spin quantum computers.
\end{abstract}


 \maketitle

Vacuum electromagnetic (EM) fluctuation plays an important role in
quantum dynamics of nuclei \cite{Purcell}, atoms
\cite{Lamb1}, electrons \cite{Electron_cavity}, and solid-state
quantum structures \cite{Yablonovitch}. Recent advances in optical
micro-structures, such as micro-spheres
\cite{microsphere2_vahala,WangHL_MCdot}, micro-rings
\cite{Vahala_toroid,waveguide_cavity}, micro-pillars
\cite{Yamamoto_QDcavityPhoton}, and engineered defects in photonic
crystals \cite{Akahane_PhotonicCrystal}, offer an opportunity of
modifying the vacuum EM environment of various systems with a
great extent of controllability. Novel ideas have been
demonstrated such as engineering the Casimir force in
microelectromechanical systems \cite{Capasso_Casimir} and
assembling semiconductor quantum dots inside photon resonators for
efficient single photon sources \cite{Yamamoto_QDcavityPhoton}. At
the same time, modern technology in ultrafast optics allows almost
arbitrary design of laser pulses for coherent control
\cite{Goswami_pulse}. Optical control of excitons in quantum dots
has become one of the methods for building solid-state
processors of quantum information \cite{Steel_QDgate,ORKKY,Chen_Raman}.

Possessing ultra-long coherence time, electron spins in quantum dots
are among the top candidates for quantum computing
\cite{Loss_QDspinQC,Imamoglu_CQED_Spin}.
Recent advances in quantum optics experiments have encouraged the
proposals of optical manipulation of spins on the picosecond
scale. In particular, Raman process \cite{Chen_Raman} and optical
RKKY interaction \cite{ORKKY} mediated by charged excitons have
been proposed for single-spin and two-spin operations,
respectively, which constitute a set of gates for universal
quantum computing. To make quantum computing complete, reading
(measurement) and writing (initialization) of the qubits are two
basic steps. Because the direct coupling between spins and their
EM environment is extremely weak (which favors the
long coherence time), spin readout has been a formidable task
under investigation
\cite{read_Martin_FETexp,read_Loss,read_Popescu,read_Kouwenhoven,
read_Friesen,read_NV_exp,read_Gammon}.
In most existing schemes
\cite{read_Martin_FETexp,read_Loss,read_Popescu,read_Kouwenhoven,read_Friesen},
the spin state is first mapped into an orbit state which is then
detected by electric sensors. These readout schemes, limited by
the clock speed of the electric measurement and/or requiring local
magnetic control,
is not ideally
suited for optically operated spin quantum computers.
The ultrafast initialization of an individual qubit is essential
in quantum error correction and fault-tolerant quantum computing
\cite{Shor96_QFT},
for the error (entropy)
has to be erased as it is generated during processing
\cite{DiVincenzo_Criteria_7}.
Much less attention has been paid to the speed of
initialization \cite{read_Friesen} than its importance would suggest.

Quantum non-demolition (QND) measurements enable scalable quantum
computing in the presence of less than ideal detector efficiency.
By contrast, ensemble measurements for
scalable computation may be problematic
\cite{liu_ensemble}.
In a large fraction of currently known quantum algorithms,
especially the seminal Shor's algorithm for factorization \cite{Shor94} and its
relatives,
while the terminal
state is a superposition of the qubit basis states $\sum_x
C_x|x\rangle$ ($x$ denotes an $N$-bit binary number), exponential
speedup over the classical algorithms depends on the information
retrieval by a single measurement resulting in the
projection into a computational basis state.
However, implementation of one-shot measurement is difficult. In
practice, a readout scheme may suffer from detection inefficiency
which may be remedied by repeating the read  cycle  to gain
sufficient accuracy. In a QND readout, a state collapses into a
certain state $|x\rangle$ after the first reading cycle and
remains in this state for the repeated cycles, so a multi-shot
measurement just gives the same result as a single-shot
measurement with 100\% efficiency. On the contrary, if the
readout is destructive, the entire algorithm has to be
rewound from the very beginning, resulting in an ensemble of
resultant states $|x\rangle$ after readout. If the ensemble
measurement is an uncorrelated one (i.e., the different qubits are
independently measured), it cannot distinguish different
superposition states, e.g., the two states
$|0000\rangle+|0011\rangle+|0110\rangle+|1001\rangle+|1100\rangle+|1111\rangle$
and $|0000\rangle+|0101\rangle+|1010\rangle+|1111\rangle$.
Or if the readout is a correlated measurement with coincidence
counting, the number of counting channels is of the order of $2^N$.

\begin{figure}[b]
\begin{center}
\includegraphics[width=3.4cm, height=2.1cm
, bb=105 555 445 765, clip=true ]{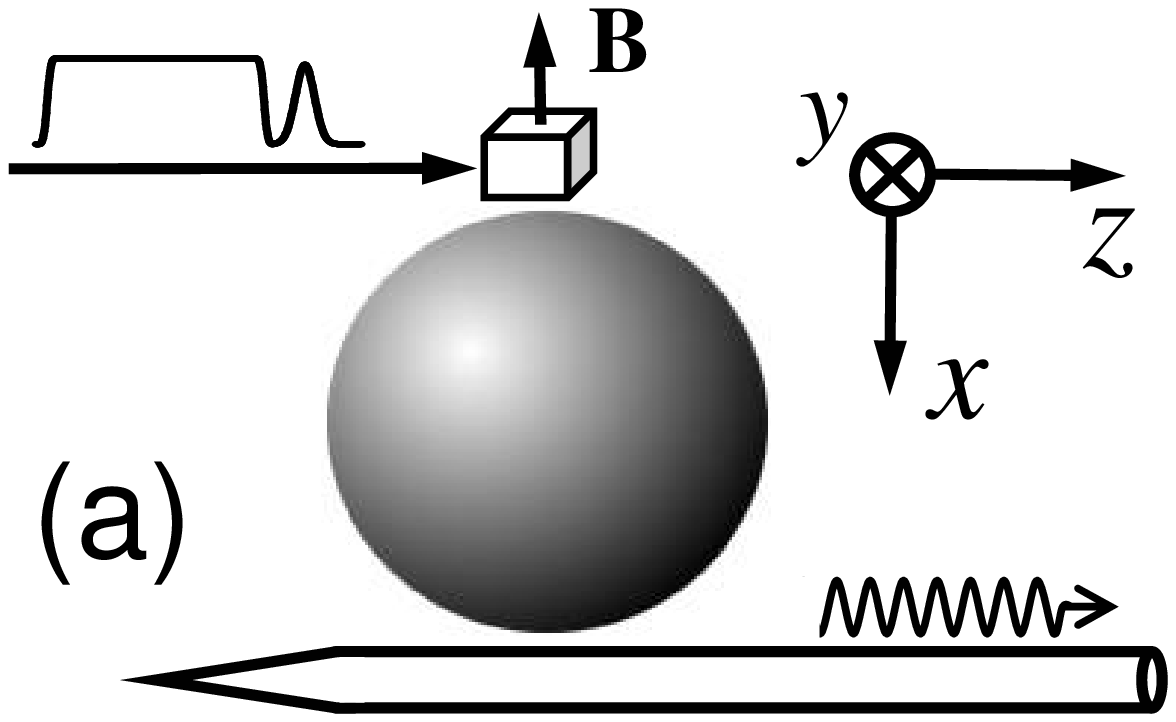} \hspace{1.5cm}
\includegraphics[width=3.05cm, height=2.1cm
, bb=128 571 433 781, clip=true ]{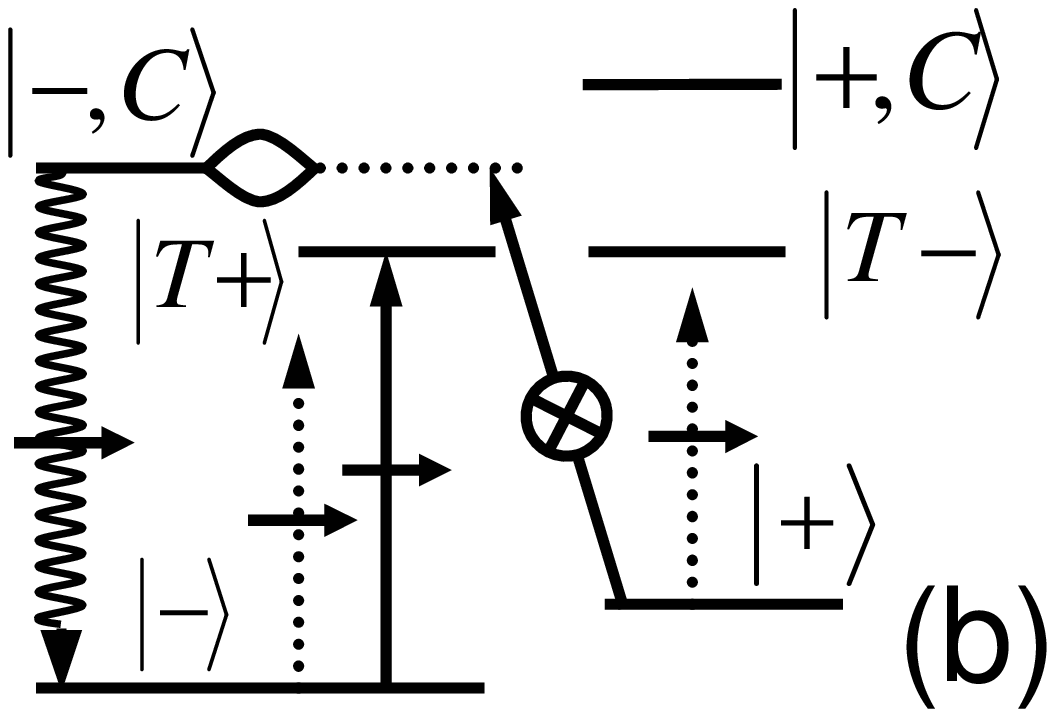}
\end{center}
\caption{(a) Physical structure of the coupling system of a doped
nanodot, a microsphere, and a tapered fiber. (b) Basic optical
processes for reading and writing a spin state. The dotted arrows
represent the AC Stark-pulses, the solid arrows with $X$ or $Y$
polarization represent the tipping pulse for writing or reading,
respectively, and the wavy arrow represents the spontaneous
emission.} \label{RWstructure}
\end{figure}

In this letter, we propose a scheme of controlling the coupling between a
nanodot and its EM environment both in space and in time to effect an
ultrafast initialization and a QND readout of the qubit represented
by an electron spin in a single nanodot.
The basic idea is depicted in Fig.~\ref{RWstructure}. A
high-quality microsphere (with Q-factor as high as $10^6\sim
10^8$) \cite{WangHL_MCdot} attached to a nanodot dramatically
modifies the EM vacuum in the vicinity of the dot via the coupling
between the evanescent wave of the resonance modes in the cavity
and the electronic transitions in the dot.
A tapered fiber coupled to the microsphere
\cite{microsphere2_vahala} acts as a
quantum channel into which a photon stored in the cavity can
escape rapidly. The spin qubit in the nanodot is controlled by the Raman processes
\cite{Chen_Raman} via the electron-trion transitions which may be
detuned far off-resonance from the cavity modes so that the quantum
channel cannot act as a source of decoherence.
When the nanodot-cavity coupling is desired,
the trion transitions and the cavity modes may be
brought  into resonance by the AC Stark
effect of a
laser light on the structure,
during which time, a tipping pulse
flips the electron state to a trion state which,
in combination with the cavity mode, relaxes back rapidly by
spontaneously emitting a photon into the quantum channel.
A choice of the polarization of the tipping pulse can
either (1)  transfer the entropy of the spin to the photon qubit in
the quantum channel, thus setting the spin qubit in
a basis state,
or (2) entangle the spin qubit with a photon qubit in the quantum
channel, thus enabling the readout of the spin qubit via the photon
detection. While optical pumping and spontaneous emission have been proposed
for spin readout and initialization \cite{read_Gammon}, our
computation suggests that the spatial-temporal engineering of
nanodot-vacuum coupling proposed here increases the efficiency by
several orders of magnitude,  leading to ultrafast initialization
and QND-readout which are suitable for, respectively,
fault-tolerant and scalable quantum computing.

\begin{figure}[t]
\begin{center}
\includegraphics[
width=6.6cm, height=8.01cm, bb=42 123 557 748, clip=true ]{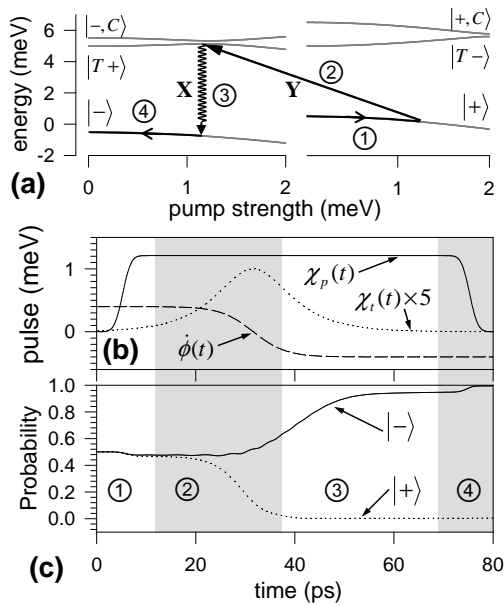}
\end{center}
\caption{(a) Detailed optical process for spin initialization.
The grey curves are the energies of different states versus the
Rabi frequency of the AC Stark pulse. The energies of the cavity
modes and the trion states are measured from the central frequency
of the AC Stark pulse.
(b) The Rabi frequencies of the AC Stark pulse and the
tipping pulse (amplified by a factor 5), and the
sweeping frequency of the tipping pulse.
(c) Probabilities of spin down and up.
Different steps of the writing cycle,
indicated by \textcircled{1}-\textcircled{4}, are distinguished by
shadowed areas in (b) and (c). The parameters used are as follows:
The Zeeman splitting of the spin $\omega_c=1$ meV, the
cavity-trion coupling $g_{\rm cav}=0.1$ meV, the cavity-fiber
coupling $\gamma=0.2$ meV, the detuning of the cavity state
$|-,C\rangle$ from the trion state $|{T+}\rangle$ is $\Delta=0.5$
meV, the damping rate of trions $\Gamma=1$ $\mu$eV, the
cavity-free space loss rate $\gamma'=0.045$ $\mu$eV (corresponding to
$Q \sim 3\times 10^7$), and the dipole matrix element of
the cavity mode is 0.3 times that of the trion state.}
\label{Writing}
\end{figure}

The detailed optical processes of writing a spin qubit are
illustrated in Fig.~\ref{Writing} (a). Without loss of generality,
we assume that the doped electron is initially in an unpolarized
state, i.e., $\hat{\rho}(-\infty)=0.5|-\rangle\langle
-|+0.5|+\rangle\langle +|$, where $|\pm\rangle$ are the two spin
states split by a strong magnetic field applied in the
$x$-direction. The two degenerate trion states $|{T\pm}\rangle$
can be excited from $|\mp\rangle$ or $|\pm\rangle$ by a $X$- or
$Y$-polarized pulses, respectively \cite{SolidstatephaseGate}. The
evanescent wave of a relevant whispering-gallery-mode in the
cavity is $X$-polarized in the vicinity of the nanodot, so that
when brought within resonance by the AC Stark pulse, the trion
states $|{T\pm}\rangle$ and the cavity states $|\mp,C\rangle$ are
coupled into two split trion-polariton states, respectively. A
writing cycle consists of four basic steps: (1) An  $X$-polarized
AC Stark pulse is adiabatically switched on, bringing the states
$|{T+}\rangle$ and $|-,C\rangle$ into resonance; (2) A tipping
pulse with $Y$-polarization flips the spin up state $|+\rangle$ to
the polariton states formed by $|{T+}\rangle$ and $|-,C\rangle$;
(3) The polariton states relax to the spin down state $|-\rangle$
rapidly by emitting a photon into the tapered fiber, dissipating
the entropy of the system to the EM environment; (4) The AC Stark
pulse is adiabatically switched off. Ideally, after the writing
cycle, the spin is fully polarized, i.e., $\hat{\rho} =
|-\rangle\langle -|$ and the entropy (or the quantum information
if the electron is initially in a pure state) of the spin is
mapped into the quantum channel.

We tested the fidelity and duration
of the writing process by
simulating the master equation of the system
with realistic parameters
given in the caption of Fig.~\ref{Writing}.
Fig.~\ref{Writing} (c)  shows that a single writing cycle
completed within 80 ps produces an almost 100\% polarized spin
from a maximally mixed state. The density matrix at the end of the
cycle is $\hat{\rho}=0.9945 |-\rangle \langle -|+0.0040|+\rangle\langle
+|+\hat{\rho}_{\rm err}$, where $\hat{\rho}_{\rm err}$ is the probability
($\approx$0.15\%) of the system remaining in the trion states. The
simulation included reasonable estimates of the
decay of the trion and the cavity modes by emitting
photons into free-space EM modes. Depending on the polarization of the emitted
free-space photon, the trion state relaxes to different spin states, which is
the main source of the writing error ($\approx 0.4\%$). The
multi-photon cavity states were included in the numerical
calculation, as they renormalize the AC Stark shift (the real
excitation of multi-photon states is negligible due to the
off-resonance condition).
Inclusion of up to 3-photon states was found sufficient to obtain converged results.

The temporal control of the nanodot-vacuum coupling
is provided by the designed shaping of the pump and
tipping pulses. The $X$-polarized AC Stark pulse has
an almost-square profile as
\begin{equation}
\chi_{p}(t)=
{\chi_p}e^{-i\Omega_pt}
\left[\text{erf}\left({\sigma_p(t-t_1)}\right)-
\text{erf}\left({\sigma_p(t-t_2)}\right)\right], \nonumber
\end{equation}
[see Fig.~\ref{Writing} (b)]. The spectral width
($\sigma_p=0.354$~meV) is set much smaller than the detuning ($\Omega_p$ is 5.5 meV
below the $|-\rangle\rightarrow|{T+}\rangle$ transition), so that
the effect due to non-adiabatic switch-on and off is negligible
(the error $<0.15\%$).
For the parameters in Fig.~\ref{Writing}, the trion state
$|{T+}\rangle$ and the cavity state $|-,C\rangle$ are brought into
resonance when the pump strength $(2\chi_p)$ reaches the value 1.21
meV. As the pump pulse maintains the resonant cavity-dot tunnelling
which facilitates the photon escape to the quantum channel, the
trion state relaxes very fast (on the time-scale of 10 ps).
A duration of the pump pulse $t_2-t_1=70$ ps is found
sufficient for the total dissipation of the photon. The tipping
pulse ideally should be a $\pi$-pulse for Rabi-oscillation between
$|+\rangle$ and $|{T+}\rangle$. Due to the dynamical nature of the
states (dressed by the AC Stark pulse) and the rather small
polariton splitting ($\sim 0.1$ meV), a perfect $\pi$-rotation
requires an extremely long pulse. The solution is to
shape a chirped pulse as $\chi_t(t)=\chi_t e^{-i\phi(t)-i\Omega_t
t} {\rm sech}\left(\sigma_t (t-t_t)\right)$ with the phase
sweeping rate  $\dot{\phi}(t) =-\sigma_c \tanh\left(\sigma_t
(t-t_t)\right)$ \cite{Goswami_pulse}. The frequency of the pulse
now will sweep from $\sigma_c$ above $\Omega_t$ to $\sigma_c$
below. When the central frequency $\Omega_t$ is tuned in between
the split polaritons and the sweeping range covers both states,
the initial spin state $|+\rangle$ will be left adiabatically in a
superposition of the two polariton states, both of which relax
rapidly to the target spin state $|-\rangle$. In simulation, the
tipping pulse, with frequency sweeping range $\sigma_c=0.4$~meV,
strength $\chi_t=0.2$~meV, and duration $1/\sigma_t=6.58$~ps,
flips the spin state $|+\rangle$ to the polariton states with
negligible error.

\begin{figure}[t]
\begin{center}
\includegraphics[width=6.6cm, height=7.53cm,
bb=40 124 559 716, clip=true]{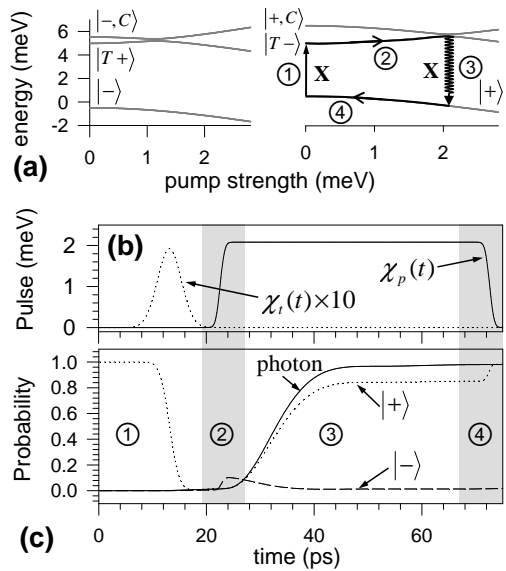}
\end{center}
\caption{(a) Detailed optical process for spin readout. (b) The
Rabi frequencies of the AC Stark pulse  and the tipping pulse (amplified by a factor 10).
(c) Probabilities of spin down,  spin up, and photon emission,
for a spin initially polarized up. Different steps of the
reading cycle, indicated by \textcircled{1}-\textcircled{4}, are
distinguished by shadowed ares in (b) and (c). The parameters are
the same as in Fig.~\ref{Writing}.} \label{Reading}
\end{figure}

A mere switch of the polarizations of the tipping and pump
pulses from $(Y,X)$ to $(X,Y)$,  respectively,
changes the ``write'' operation to a ``read'' one.  The reading cycle, as
illustrated by Fig.~\ref{Reading} (a), includes four basic steps:
(1) An $X$-polarized tipping pulse flips the spin state
$|+\rangle$ to the trion state $|{T-}\rangle$; (2) A $Y$-polarized
AC Stark pulse  adiabatically switched on drives the trion state
into resonance with the cavity state $|+,C\rangle$; (3) The trion
state resonantly tunnels into the cavity state and relaxes rapidly
back to the spin state $|+\rangle$, leaving a photon emitted into
the quantum channel; (4) The AC Stark pulse is adiabatically
switched off. Suppose that the spin state to be read is
$\alpha|+\rangle+\beta|-\rangle$ and the channel is initially in
the vacuum state $|0\rangle$. The read process will ideally
transform the system into the entangled state
$\alpha|+\rangle|1\rangle+\beta|-\rangle|0\rangle$ (where
$|1\rangle$ denotes the photon wave packet emitted into the
fiber),
so the spin state can be read out by measuring the
photon number state with a photon detector. The photon
counting doesn't disturb a spin eigenstate,
providing a QND measurement of the spin.

Note that the pulse timing for reading is different from that for
writing [see Fig.~\ref{Reading} (b)].
The reading sequence has been designed to minimize the real
excitation of the multi-photon states, while the writing sequence
has been designed to minimize the emission of free-space photons
by the trion states.
In reading, the Rabi flip process is between steady states and the transition
is well separated from the others, the pulse-shaping trick is
unnecessary, so the tipping pulse is assumed the simple Gaussian
form $\chi_t(t)=\chi_te^{-\sigma_t^2(t-t_t)^2/2-i\Omega_t t}$ with
optical area equal $\pi$. The AC Stark pulse is chosen
$Y$-polarized to avoid direct excitation of
the cavity mode.

The reading cycle has been numerically simulated
for the same structure as in Fig.~\ref{Writing}. The tipping and
the AC Stark pulses are set such that $1/\sigma_t=2.19$~ps,
$\chi_t=0.192$~meV, $\Omega_t$ is in resonant with the
$|+\rangle\rightarrow |{T-}\rangle$ transition, $\sigma_p=0.707$~meV,
$2\chi_p=2.08$~meV, $\Omega_p$ is 5.5~meV below the
$|-\rangle\rightarrow |{T-}\rangle$ transition, and the duration of
the pump pulse is $t_2-t_1=50$~ps. After a single cycle of
reading, an initial state $\hat{\rho}_0=|+\rangle \langle +|$ results in
the final state
$
\hat{\rho}_1= 0.0161|-\rangle \langle
-|+0.9824|+\rangle\langle +|+\hat{\rho}_{\rm err} $
with probability
0.9806 of having a photon emitted into the quantum channel [see
Fig.~\ref{Reading} (c)], while an initial state $\hat{\rho}_0=|-\rangle
\langle -|$ results in the final state $\hat{\rho}_1=0.9955 |-\rangle
\langle -|+0.0040|+\rangle\langle +|+\hat{\rho}_{\rm err}$ with
probability 0.0015 of having a photon emitted into the quantum
channel (not shown). The photon emitted into the fiber can be
detected with high efficiency \cite{Yamamoto_detector}. If the
detector has zero dark-count rate and efficiency of 50\%, the POVM
operators for the reading process  are  defined as $\hat{P}_-\equiv
0.9992|-\rangle\langle-|+0.5097 |+\rangle\langle+|$ and $\hat{P}_+\equiv
0.0008|-\rangle\langle-|+0.4903 |+\rangle\langle+|$. Within 5
reading cycles, the spin state can be read out with accuracy
higher than 97\%, and the back-action noise to the spin is less than
10\%, while the time duration is less than
0.4 ns, much shorter than the spin decoherence time.

In summary, the coupling between the EM vacuum and nanodots can be
customized both by spatially assembling micro-resonators and
quantum channels in the vicinity of the dot and by
temporal design of  the control optical pulses.
Such a control can be employed for ultrafast initialization and QND readout of a single
electron spin in a doped nanodot. The readout and initialization
schemes, proposed here for a specific fiber-microsphere structure,
can be implemented in alternative systems such as
etched waveguide-resonator structures on semiconductor surfaces
\cite{Vahala_toroid,waveguide_cavity} and line- and point-defects
engineered in photonic crystals \cite{Akahane_PhotonicCrystal}. Though high-$Q$
cavity has been assumed in numerical simulations of the initialization and readout
operations, efficiency reduction resulting from lower $Q$ values can be tolerated
by recycling the operations for a few times.
The schemes proposed here may also be adapted to monitor and control
the spin state of a single molecule adsorbed to a cavity-fiber
structure \cite{read_singleMolSpin}.

This work was supported by  ARDA/ARO DAAD19-02-1-0183, NSF
DMR-0099572,  and QuIST/AFOSR F49620-01-1-0497.


\end{document}